# The network structure of scientific revolutions

**Authors:** Harang Ju[1], Dale Zhou[1], Ann S. Blevins[2], David M. Lydon-Staley[2,3], Judith Kaplan[4,5], Julio R. Tuma[5,6], Danielle S. Bassett[2,7-11]*.

**Affiliations:** [1]Neuroscience Graduate Group, University of Pennsylvania, Philadelphia, PA 19104, USA; [2]Department of Bioengineering, School of Engineering & Applied Science, University of Pennsylvania, Philadelphia, PA 19104, USA; [3]Annenberg School for Communication, University of Pennsylvania, Philadelphia, PA 19104, USA; [4]Department of History and Sociology of Science, College of Arts and Sciences, University of Pennsylvania, Philadelphia, PA 19104, USA; [5]Integrated Studies Program, College of Arts and Sciences, University of Pennsylvania, Philadelphia, PA 19104, USA; [6]Department of Philosophy, College of Arts and Sciences, University of Pennsylvania, Philadelphia, PA 19104, USA; [7]Department of Electrical & Systems Engineering, School of Engineering & Applied Science, University of Pennsylvania, Philadelphia, PA 19104, USA; [8]Department of Physics & Astronomy, College of Arts & Sciences, University of Pennsylvania, Philadelphia, PA 19104, USA; [9]Department of Neurology, Perelman School of Medicine, University of Pennsylvania, Philadelphia, PA 19104, USA; [10]Department of Psychiatry, Perelman School of Medicine, University of Pennsylvania, Philadelphia, PA 19104, USA; [11]Santa Fe Institute, Santa Fe, NM 87501, USA; *Correspondence to: dsb@seas.upenn.edu.

**Abstract:** Philosophers of science have long postulated how collective scientific knowledge grows. Empirical validation has been challenging due to limitations in collecting and systematizing large historical records. Here, we capitalize on the largest online encyclopedia to formulate knowledge as growing networks of articles and their hyperlinked inter-relations. We demonstrate that concept networks grow not by expanding from their core but rather by creating and filling knowledge gaps, a process which produces discoveries that are more frequently awarded Nobel prizes than others. Moreover, we operationalize paradigms as network modules to reveal a temporal signature in structural stability across scientific subjects. In a network formulation of scientific discovery, data-driven conditions underlying breakthroughs depend just as much on identifying uncharted gaps as on advancing solutions within scientific communities.

**One Sentence Summary:** The authors test theories of scientific progress on growing concept networks and reveal data-driven conditions underlying breakthroughs.

**Main Text:** Philosophers of science have long postulated a range of processes underlying scientific progress. In 1959, Popper described the development of scientific ideas as a sequence in which previous theories are falsified (*1*), and in 1962, Kuhn as periods of *normal science*, in which researchers "solved puzzles" within a paradigm, separated by *paradigm shifts* that overturn the existing paradigm (*2*). While Lakatos in 1970 balanced the two theories by suggesting a *research programme* in which knowledge expands from a common hard core set of theoretical commitments and practices (*3*), Feyerabend in 1975 discounted any single mechanism for scientific progress (*4*). Recently, the field of *science of science* has begun to use a more quantitative approach to probe the conditions underlying scientific discovery (*5*).

Despite the variety of theories regarding scientific progress, such as cultural accounts of scientific practice (*6*), many suggest a dependence of new discoveries on the existing body of knowledge. Newton writes in 1675, "If I have seen further, it is by standing on the shoulders of giants" (*7*). Indeed, discoveries, including calculus, are often *multiples*, discovered independently and contemporaneously (*8*). These observations prompt the question of how an existing body of knowledge influences the discovery of new scientific knowledge.

Here, we expand upon the concept of a body of knowledge, not as amorphous, but as comprised of distinct relationships between concepts. We formalize this structured body of knowledge as a *concept network* whose nodes represent concepts and whose edges represent inter-concept relations (*9*). To study the process of science, we build growing concept networks from Wikipedia, a free online encyclopedia (**Fig. 1A**; see Supplementary Materials and Text for details). Each Wikipedia article explains a concept and contains hyperlinks to other Wikipedia articles, forming a directed network of concepts. For each article we create a node, and for each hyperlink from the lead section (i.e., the introduction) of an article we create a directed edge from the *hyperlinked* article to the *hyperlinking* article. We weight the directed edges to represent the similarity between two concepts, quantified using the cosine similarity between two articles' vector representations that are derived from *tf-idf* (i.e., term frequency-inverse document frequency) encoding. Focusing on the article's history and lead sections, we parse the earliest year associated with the concept's discovery and set that year as an attribute for each node. We use this information about the year of a concept's discovery to create a growing network across time, adding edges at the same time as the addition of their nodes.

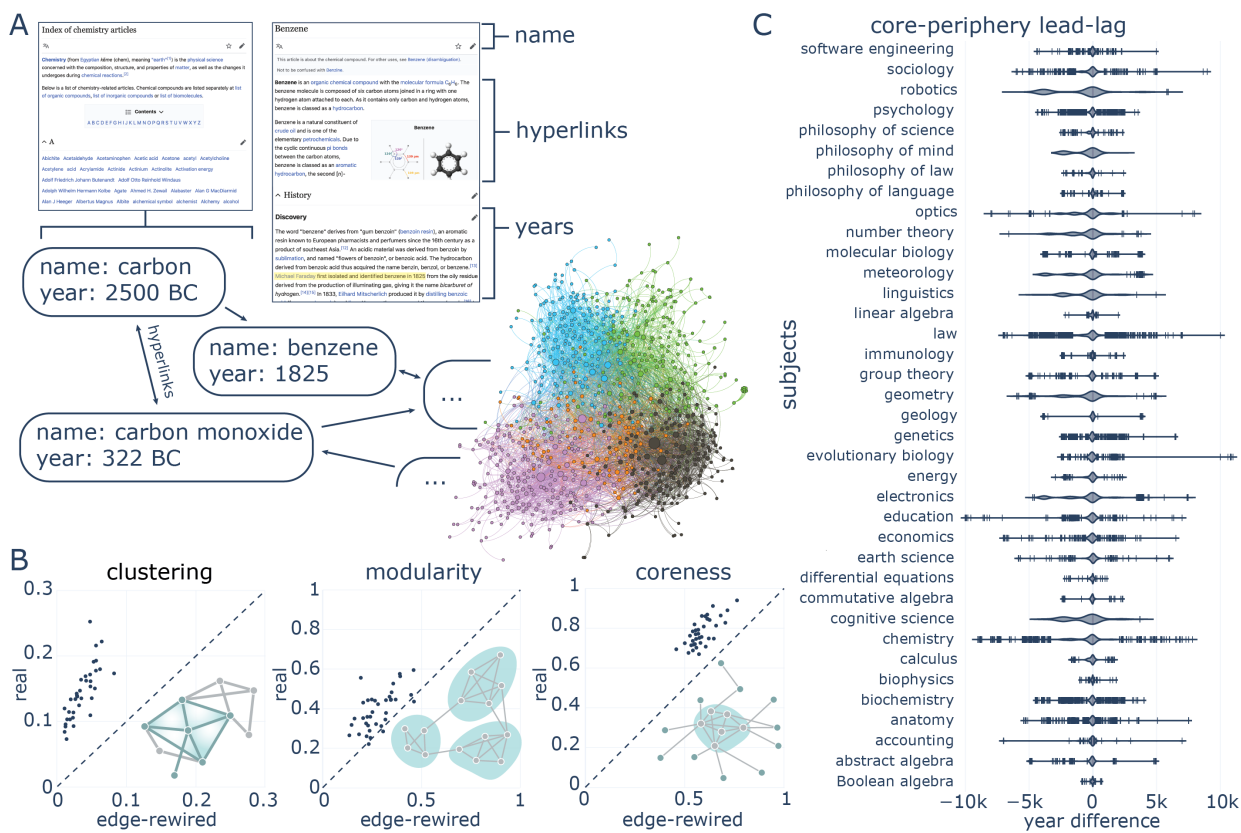

**Fig. 1. Building a growing concept network from Wikipedia.** (A) Each network is made up of Wikipedia articles that are indexed by subject in the natural sciences, mathematics, and the social sciences. Each node corresponds to an article; the node's name is the title of the article, and the node's birth year is the first year listed in the introduction or history sections as the year when the concept was conceived (highlighted in yellow). Each directed edge corresponds to a hyperlink, from the article that is hyperlinked to the article that hyperlinks. (B) The final concept networks display greater clustering, modularity, and coreness than null networks constructed to reflect Fereyabend's hypothesis that science lacks a characteristic pattern of discovery. (C) By comparing the birth year of core nodes to those of neighboring peripheral nodes, we observed that there is no clear *lead-lag* relationship between a core node and its neighboring peripheral

nodes. Violin plots show distributions of the years of core nodes minus the years of neighboring peripheral nodes, for each core-periphery edge. Vertical lines indicate outliers, which are less than $Q_1 - (1.5 \times IQR)$ or greater than $Q_3 + (1.5 \times IQR)$ where $Q_n$ is the $n^{th}$ quartile and $IQR = (Q_3 - Q_1)$. Negative values indicate that core nodes are discovered before their neighboring peripheral nodes, and *vice versa*.

Given the notions of characteristic patterns in discovery by Kuhn and Lakatos, or the lack thereof by Feyerabend, we first quantified the structure of concept networks using network science. As an initial test of Kuhn's hypothesis of *normal science*, we measured whether concepts forms clusters, in which scientists "solve puzzles" within an existing paradigm, operationalized at the node level using the clustering coefficient, which quantifies the local density of connections (*10*), and at the network level using modularity, in which nodes are densely connected within modules and loosely connected across modules (*11*). Additionally, we used the core-periphery measure (*12*) to measure whether networks form a "common core" as suggested by Lakatos's *research programme*, where *core* nodes are densely connected to each other and *peripheral* nodes are loosely connected to the core. We compared these measures in real networks to those in null networks that destroy existing patterns in network structure through random rewiring, thereby operationalizing Fereyabend's hypothesis of a lack of pattern to discovery. We found that real networks have greater clustering coefficients, modularity, and coreness than those in the edge-rewired networks (**Fig. 1B**). These findings suggest structure in concept networks, in which concepts are clustered with modules (akin to subfields) and characterized by a core set of concepts.

Given the structure of concept networks at the current time, we wished to examine any patterns in the evolution of those structures over the course of history. By comparing the birth year of core nodes to the birth year of neighboring peripheral nodes, we observed that there is no clear lead-lag relationship between a core node and its neighboring peripheral nodes (**Fig. 1C**). That is, core nodes do not necessarily precede their neighboring peripheral nodes in their discovery, and the same is true for core-periphery structures within specific modules (**Fig. S1**). Interestingly, a few core nodes are born consistently earlier than most peripheral nodes. These nodes, such as the node "Hydrology" in subject "Earth Science", may serve as concepts that are central to the subject as a whole but comprise a "hard inner" part of the core (**Fig. S2**). Taken together, these results point to both an outward expansion and an inward exploration of concepts, in which the core of a research programme is often updated by new discoveries that are influenced by discoveries that occur in the periphery.

How then does a body of knowledge grow? Because real concept networks are highly clustered with both inward and outward growth, we hypothesized that concept networks fill gaps in knowledge (*13*), akin to Kuhn's puzzle-solving *normal science*. To test their relevance to discovery, we formalize knowledge gaps in the language of algebraic topology (*14*), where a gap corresponds to a topological cavity in any dimension *n* (**Fig. 2A-B**; see Supplemental Methods for details) (*15*). To detect gaps within the growing concept network, we use persistent homology, which chronicles the birth, evolution, and collapse of topological cavities across a growth process (*16*). In general, a cavity in a Wikipedia network is a set of articles that are connected to each other, but no article in the set connects all articles of the cavity.

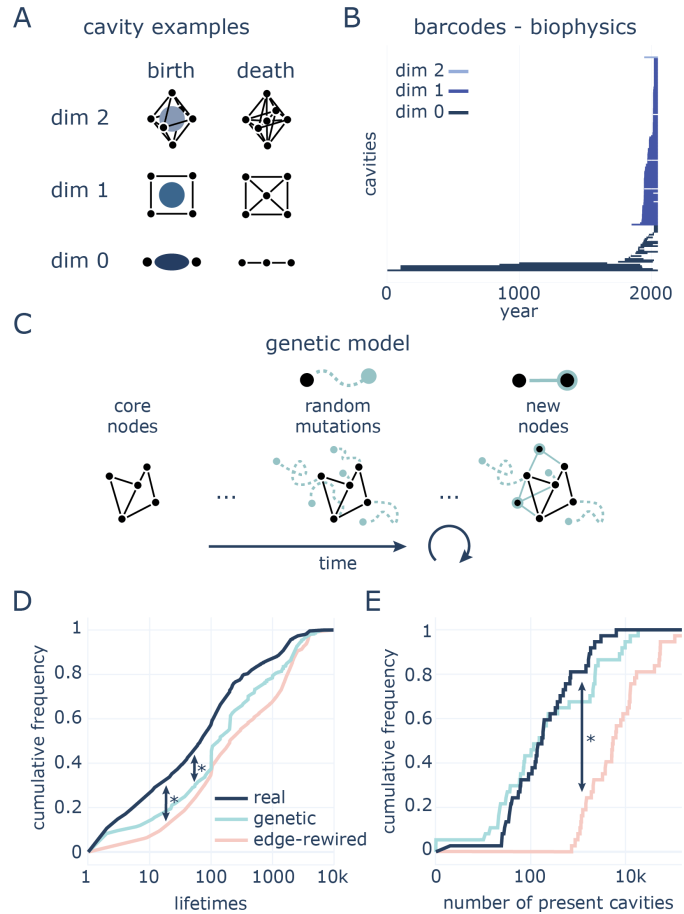

**Fig. 2. Real concept networks maintain shorter and fewer gaps.** (A) Examples of cavities in 0, 1, and 2 dimensions. Cavities are born when a new, added node creates a topological gap; cavities die when the gap is closed by a new node and its edges tessellate the gap. (B) Barcode for the *biophysics* network. The left and right points of each bar are the birth and death times of a persistent cavity. (C) Illustration of genetic model for concept network growth and evolution. (D) Real networks have knowledge gaps for shorter duration than either random ($KS = 0.24$, $p = 1.4 \times 10^{-188}$) or genetic model ($KS = 0.20$, $p = 1.3 \times 10^{-15}$) networks. (E) Real networks have fewer knowledge gaps that are currently present (i.e., that have yet to die) than random networks ($KS = 0.81, p = 2.1 \times 10^{-12}$).

To further examine the process of scientific discovery, we first formulated a genetic null model, the network structure of which we can compare to that of a real concept network. The genetic model simulates scientists learning about existing concepts and slowly mutating them to create new concepts but without a "preference" (i.e., an objective or fitness function) for how new concepts are mutated (**Fig. 2C**). More formally, the model is initialized as a subset of a real network, containing "core nodes" that were born before a predetermined year (BC 500 in our simulations). For each node, we iteratively mutate a *tf-idf* vector representation of the node's Wikipedia article. We then create a new node from the mutated vector and connect it to similar nodes in the network (see Supplemental Methods and **Fig. S3**).

To test our hypotheses about the presence of gap-filling and its contribution to discovery, we computed the persistent homology of real networks (**Fig. 2B**). For persistent cavities that have already died by $t_{max}$, those in real networks have significantly shorter lifetimes than those in either random (Kolmogorov-Smirnov statistic, $KS = 0.24$, $p = 1.4 \times 10^{-188}$) or genetic model ($KS = 0.20$, $p = 1.3 \times 10^{-15}$) networks (**Fig. 2D**). Further, real networks have significantly fewer persistent cavities that are still present at $t_{max}$ than random networks ($KS = 0.81$, $p = 2.1 \times 10^{-12}$; **Fig. 2E**) but not genetic networks. Collectively, these results support our hypothesis that real concept networks fill gaps more quickly than null networks and leave fewer gaps alive at the present day.

In addition to the first part of Kuhn's theory regarding puzzle-solving, which we operationalized as cavity-filling, the second part involves paradigm shifts that radically change the way scientists view concepts. In our network formulation, we operationalized paradigms as temporal modularity structure, where an "incommensurable" paradigm shift would drastically change the membership of nodes to modules. We built a multilayer network with each layer containing the concept network at each year (*17*) and used community detection to identify modules across time (*18*). Then, we calculated the number of times that a node changes its membership to a module for each year (**Fig. 3A**) and performed change point detection on the number of changes to identify epochs of stability in module membership (**Fig. 3B**; Supplemental Methods) (*19*). We found that after an initial, short epoch of little change, the network enters an epoch of moderate and lasting change (**Fig. 3C**). Then, the network enters a short epoch of much greater change, after which the network stabilizes in the last epoch. This signature breaks in random networks (**Fig. S4**). The data demonstrate that shifts in the structure of concepts occur not as abrupt Kuhnian shifts but as gradual Lakatosian modifications (*20*) and that subjects share a signature in structural stability across time.

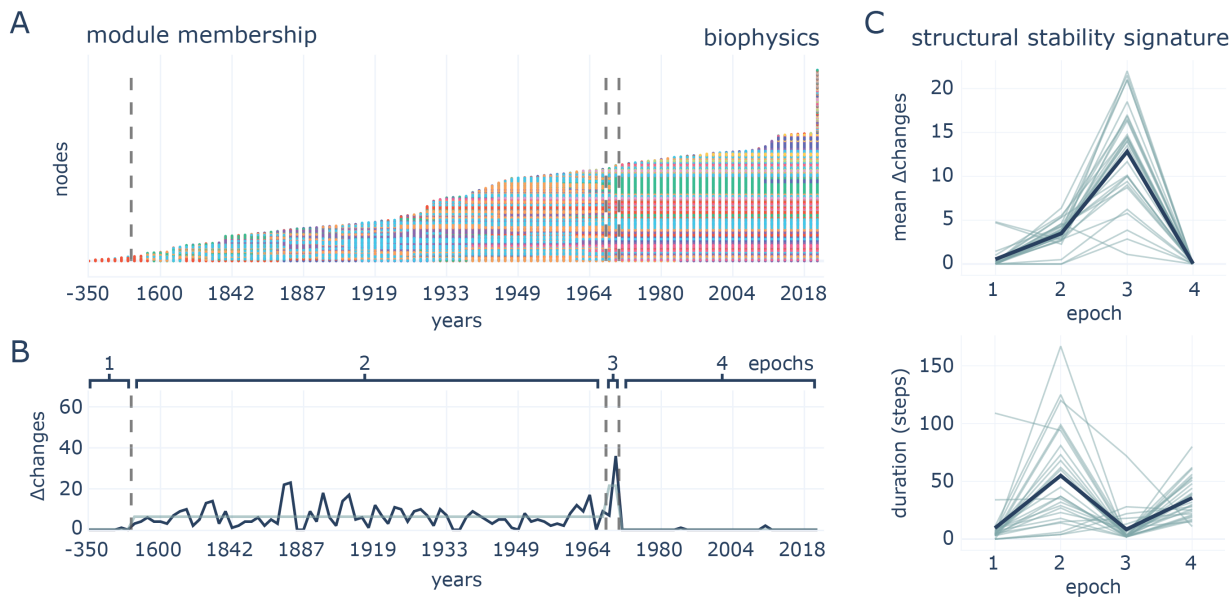

**Fig. 3. Concept networks undergo a signature pattern in structural stability.** (A) Module membership (colored) of nodes across years for the *biophysics* network. (B) The number of changes in module membership for the *biophysics* network. Dashed lines are changepoints in epochs, and the teal line is the mean for each epoch. (C) The mean number of changes within an epoch (top) and the duration of each epoch (bottom) reveals a signature (dark green) averaged across subjects (teal).

Finally, we ask whether we can predict the merit of a discovery by measuring the node's theoretical ability to influence a body of knowledge due to its location within the gappy topology. We operationalize merit using a network-based measure and a separate, external measure: the receipt of a Nobel prize (*21*). After constructing a single network containing all nodes from all subjects, we defined a network-based measure of a node's potential influence as its impulse response, a dynamical-systems quantification of how much a network "responds" to an "impulse" that perturbs one node (**Fig. 4A**; Supplemental Methods). We observed that nodes that more frequently participate in the birth or the death of cavities have higher impulse responses (birth: Pearson's correlation coefficient $r = 0.36$, $p = 0.0$; death: $r = 0.38$, $p \ll 0.001$; **Fig. 4B**). Importantly, such nodes are also more frequently awarded Nobel prizes (birth: $KS = 0.15$, $p = 4.2 \times 10^{-6}$; death: $KS = 0.17$, $p = 1.2 \times 10^{-7}$; **Fig. 4C-D**). These results suggest that the network structure reflects the real-world influence of concepts on the body of knowledge.

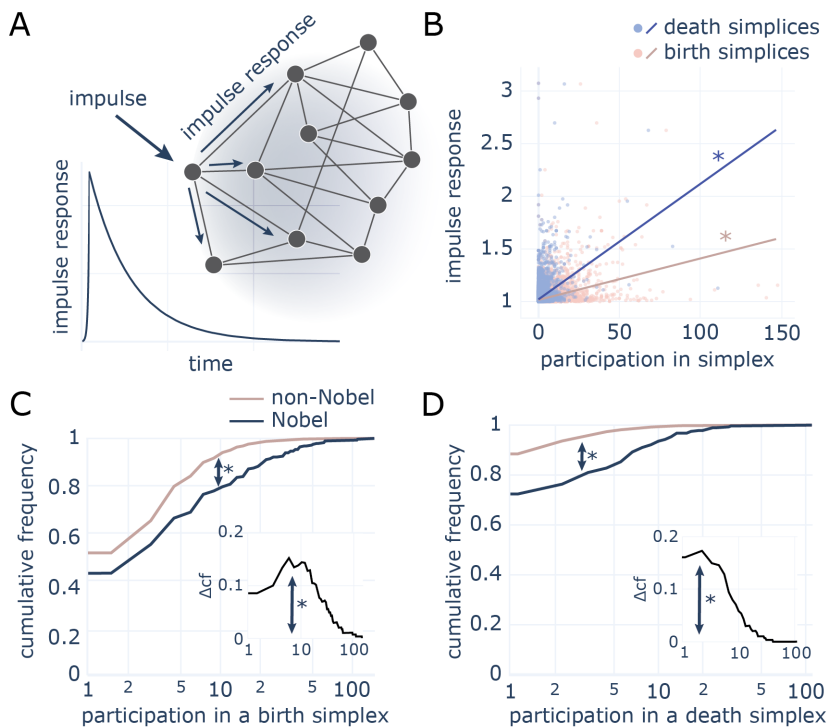

**Fig. 4. Network topology reveals the impact of concepts.** (A) Illustration of the response of a network to impulse on a node. (B) Nodes that more frequently participate in the birth or death of persistent cavities have higher impulse response, which is a dynamical-systems measure of a node's influence on a network (birth: $r = 0.36$, $p = 0$; death: $r = 0.38$, $p = 0$). (C-D) Nobel prizes are more frequently awarded for nodes that participate in the birth (panel C) and death (panel D) of persistent cavities (birth: $KS = 0.15$, $p = 4.2 \times 10^{-6}$; death: $KS = 0.17$, $p = 1.2 \times 10^{-7}$). Subpanels show the difference in cumulative frequencies ($\Delta$cf).

The findings described herein reveal that human knowledge grows by filling gaps in knowledge, perhaps driven by the collective curiosity of individual scientists (*8, 22*), through inward and outward exploration and gradual modifications to network structure. Moreover, knowledge discovered while creating and filling knowledge gaps is likely to be more influential and more frequently awarded in the scientific community. Our mathematical formulations of historical data pave the way to describe, understand, and even potentially guide scientific progress for individuals and funding agencies (*23*). Furthermore, our findings provide a data-driven approach to identifying novel contributions, especially those by underrepresented groups whose works are typically devalued yet are vital for vibrant scientific innovation (*24, 25*).

**Acknowledgments: Funding:** The authors acknowledge support from the John D. and Catherine T. MacArthur Foundation, the Alfred P. Sloan Foundation, the ISI Foundation, the Paul Allen Foundation, the Army Research Laboratory (W911NF-10-2-0022), the Army Research Office (Bassett-W911NF-14-1-0679, Grafton-W911NF-16-1-0474, DCIST-W911NF-17-2-0181), the Office of Naval Research, the National Institute of Mental Health (2-R01-DC-009209-11, R01-MH112847, R01-MH107235,

**Acknowledgments: Funding:** The authors acknowledge support from the John D. and Catherine T. MacArthur Foundation, the Alfred P. Sloan Foundation, the ISI Foundation, the Paul Allen Foundation, the Army Research Laboratory (W911NF-10-2-0022), the Army Research Office (Bassett-W911NF-14-1-0679, Grafton-W911NF-16-1-0474, DCIST-W911NF-17-2-0181), the Office of Naval Research, the National Institute of Mental Health (2-R01-DC-009209-11, R01-MH112847, R01-MH107235, R21-MH-106799), the National Institute of Child Health and Human Development (1R01-HD086888-01), National Institute of Neurological Disorders and Stroke (R01-NS099348), the National Science Foundation (BCS-1441502, BCS-1430087, NSF PHY-1554488 and BCS-1631550), and the National Institute on Drug Abuse (K01DA047417); **Author contributions:** Conceptualization, H.J., D.Z.; Methodology, H.J., D.Z., A.S.B.; Software, H.J.; Validation, H.J.; Formal analysis, H.J.; Investigation, H.J.; Resources, H.J., D.S.B.; Data curation, H.J.; Writing – Original Draft, H.J.; Writing – Review & Editing, H.J., D.Z., A.S.B., D.M.L.-S., J.K., J.R.T., D.S.B; Visualization, H.J., D.Z., A.S.B., D.S.B.; Supervision, H.J., D.S.B.; Project administration, H.J., D.S.B.; Funding acquisition, D.S.B.; **Competing interests:** Authors declare no competing interests. and **Data and materials availability:** All code is available on https://github.com/harangju/wikinet. All data used in the study are publicly available on https://dumps.wikimedia.org/enwiki.

# Supplementary Materials for

The network structure of scientific revolutions

Harang Ju, Dale Zhou, Ann S. Blevins, David M. Lydon-Staley, Judith Kaplan, Julio R. Tuma, Danielle S. Bassett

Correspondence to: dsb@seas.upenn.edu

**Supplementary materials include:**
    Materials and Methods
    Supplementary Text
    Figs. S1 to S9
    Table S1 to S3

**Materials and Methods**

Data and code availability

All data to reproduce this study are publicly available at https://dumps.wikimedia.org/enwiki. Only two files are required for reproduction: (1) enwiki-DATE-pages-articles-multistream.xml.bz2 and (2) enwiki-DATE-pages-articles-multistream-index.txt.bz2, where DATE is the date of the dump. Both files are multistreamed versions of the zipped files, which allow the user to access an article without unpacking the whole file. In this study, we used the archived zipped file from August 1, 2019. All code to reproduce this study is publicly available at https://github.com/harangju/wikinet.

Building growing concept networks from Wikipedia

*Software package for representing networks.* We used the Python software package *networkx* (version 2.5) for most of our network representation and analysis. We used the Python package *igraph* (version 0.8.2) for representing networks for temporal module detection (see the later section "Temporal modularity structure").

*Selecting articles for a subject.* To build a concept network for a subject, we must first select Wikipedia articles that belong in a subject. Doing so in a principled manner can be difficult because there are no inherent delineations between articles of different subjects. Fortunately, Wikipedia provides indices of subjects, which list articles of a particular subject (https://en.wikipedia.org/wiki/Wikipedia:Contents/Indices). We chose to explore subjects in the subject areas of Mathematics and Logic, Natural and Physical Sciences, and Subdisciplines of Philosophy. For each subject, we built a network where nodes represented articles that are listed on the subject's index, and where edges represented the articles' hyperlinked connections.

*Connecting articles (network nodes) via hyperlinks.* After gathering a list of articles to include in a subject-specific network, we first create a node for each article. Then to connect the nodes, we select hyperlinks that are in the lead section of an article (which we will call article *A* for illustration; https://en.wikipedia.org/wiki/Wikipedia:Manual_of_Style/Lead_section). We take the hyperlinks from the lead section (i.e., the introduction) to (i) capture a concise overview of the concept, (ii) maintain a normal distribution of the number of hyperlinks, (iii) reduce edge density, and (iv) avoid spurious hyperlinks to tangentially related pages. If the hyperlinks point to other articles in the subject, we then create a directed edge from those other articles to article *A*. We chose to direct edges from the *hyperlinked* article to the *hyperlinking* article because the *hyperlinked* article is used to explain the *hyperlinking* article and thus influences the information presented in the *hyperlinking* article. **Table S1** and **S2** summarize network measures.

*Weighting network edges.* We determine the weight of each edge between two articles by calculating the cosine similarity between the vector of term frequency-inverse document frequencies (tf-idf) for words in one article and the tf-idf vector for words in the other article (*26*). We compute tf-idf by multiplying a local component (term frequency) with a global component (inverse document frequency). The measure is defined as follows:

$$tfidf_{i,j} = frequency_{i,j} \times \log_2 \frac{D}{document\_frequency_i},$$

where for term *i* in document *j*, $frequency_{i,j}$ is the number of times that term *i* occurs in document *j*, *D* is the number of documents, and $document\_frequency_i$ is the number of documents that contain term *i*. The tf-idf is a product of a token's frequency and the token's inverse document frequency. Thus, common tokens appearing very frequently in the corpus will be down-weighted while rate terms will be associated with a relatively large number. To account for differences in document length, we applied a common normalization such that the Euclidean norm of the tf-idf vector for a document became 1. After calculating the normalized tf-idf for each token, we quantified the similarity between pairs of nodes by computing the cosine similarity between pairs of vectors. The cosine similarity results in a quantification of node similarity ranging from 0 to 1; higher values indicate greater similarity of the text between two Wikipedia pages. We use all articles in Wikipedia as the corpus for the calculation. We use the Python package *gensim* to compute tf-idf.

*Denoting the birth year of a node.* We parse the years from the lead section and from a history section if the article has one. We denote the earliest year as the year when the node was "born" or conceived. There exist articles that do not have years listed in either the lead or the history section. To assign years to these articles, we first select all nodes without years whose parents (i.e., nodes with edges that link to a node) have years. For each such node, we denote its year as the year after the latest year of its parents. Then, we do the same for the remaining nodes without years. If a node still does not have a year, then we denote its year as 2020. Our results that use the year are robust to slight changes in the year (see Supplemental Text; **Fig. S9**).

*Parsing years.* To parse the years from the text, we use *regex* to identify numbers that are preceded by months (e.g., "January"), prepositions of time (e.g., "around"), conjugations (i.e., "and"), articles (i.e., "the"), and other time-related words (i.e., "early", "mid", "late") and followed by the words BC, BCE, or MYA. We also parse centuries (e.g., "19th Century") and convert them into years (e.g., 1800). We apply a negative sign to all years or centuries followed by BC or BCE for convenience in analysis, such that 1600 BC would become -1600. For Python implementation, see the function `filter_years(text)` in `module/wiki.py` in the code repository.

Null networks

We use an *edge-rewired* null networks to compare to real networks as a model for Feyerabend's hypothesis of anarchical scientific progress. For an edge-rewired network, we take each edge and randomly select a new target once; by the term *target*, we mean the node that the edge points to. Here, we are randomizing the target, similar to random networks in Maslov-Sneppen random networks (*27*), but without swapping the targets between pairs of nodes. In the context of Wikipedia articles, since we create an edge from a *hyperlinked* article to a *hyperlinking* article, we are maintaining the connection of an edge from the *hyperlinked* article but changing the connection of the edge to the *hyperlinking* article. Thus, we maintain degree distributions (**Fig. S5**) while removing patterns in network topology.

Persistent homology

In our study, we hypothesized that processes of scientific discovery create and fill gaps in concept networks. A tool from applied algebraic topology, called persistent homology, provides a well-defined formulation of such gaps as persistent topological cavities that evolve as a network grows (*14*). To calculate persistent homology for a growing network, we first create a correspondence between *k*-cliques, which are all-to-all connected subnetworks of *k* nodes, and *(k-1)*-simplices. Examples of simplices are: a node as a 0-dimensional simplex, an edge as a 1-dimensional simplex, a 3-clique as a 2-dimensional simplex, and so on for higher dimensions. Using the Python package `dionysus2`, we add each clique as a simplex into a *filtration* at the latest year in the clique. A *filtration* is formally a nested sequence of subspaces and can be thought of as a growing sequence of simplices. We may add more than one simplex to the filtration at each year, resulting in cavities with a lifetime of zero, which we remove for downstream analyses.

Finally, we use the package `dionysus2` to compute a reduced matrix which defines the indices of the start and end of cavities.

Simulations of knowledge discovery

To examine the process of scientific discovery itself, rather than just the resultant structure of a concept network, we formulated a genetic model of knowledge discovery. A model of the discovery process is important for testing our hypothesis that the body of knowledge grows by creating and filling cavities. The intuition for the model is that it simulates scientists who learn about existing concepts and slowly mutate them to form new concepts over the course of history. Importantly, the model has no "preference" (i.e., an objective or fitness function) that selects for certain features of new concepts. Thus, by comparing real networks to networks produced by the genetic model, we aim to explicitly test for whether real networks have a "preference" to fill knowledge gaps.

In simulations of knowledge discovery, we start with a subgraph of a subject network, consisting of nodes whose birth years are before 1 AD. For each subsequent year, we execute the following series of steps until either the year 2200 or the number of nodes of the model reaches that of the real subject-network: (i) initialize seeds for new nodes, (ii) mutate seeds, (iii) create new nodes for seeds, and (iv) connect new nodes to the model network. We capped the simulation at year 2200 to halt the program in the case that the model takes a long time to, if not never, reach the number of nodes in the real subject-network.

*Seed initialization.* Before we begin to mutate a concept-node, we must first obtain a vector representation of a node to mutate. Thus, for all new nodes in the model network, we initialize a "seed", which is a copy of the tf-idf vector of the "parent", which is a node in the model network. As noted in the supplemental section "Building growing concept networks from Wikipedia", the tf-idf vector of a node is the term-frequency inverse-document-frequency representation of the Wikipedia article corresponding to the node. Hence, by initializing and mutating the "seed", we are slightly modifying a vector representation of a Wikipedia article with each iteration.

*Seed mutation.* With *seed mutations*, we model scientists researching a concept by taking a previously known concept and slightly modifying it to, for example, test a hypothesis about the known concept. Thus, to iteratively mutate a seed, we take three steps for each seed and for each year of the simulation: a point mutation, an insertion, and a deletion. Each step has a probability of occurrence, and we base those probabilities on the statistics that we find in real networks: a process that we explain in detail below.

First, a point mutation swaps the value of a randomly chosen element in the *seed* vector with a new value for each year with probability $p$. The new value is drawn from the distribution of tf-idf values in the original subject network (**Fig. S3**). We set the probability $p$ to approximate the change in tf-idf vectors over years in the real network. To approximate this change in tf-idf vectors, we first compute for each edge and its two nodes (i) the absolute difference in years, which we call the *year-diff*, (ii) the sum of the absolute difference between tf-idf values, which we call the *sum-abs-diff*, and (iii) the average absolute difference between $10^5$ values randomly drawn from the original distribution of tf-idf values, which we call *avg-abs-diff*. Interestingly, there is a strong and statistically significant correlation between (i) *year-diff* and (ii) *sum-abs-diff* (**Fig. S3E**). This correlation suggests that the longer the time between the discoveries of two neighboring nodes, the more different the nodes are in their tf-idf representations. We thus approximate the probability of a point mutation $p$ as the slope of the linear regression between (i) *year-diff* and (ii) *sum-abs-diff*, normalized by dividing by (iii) *avg-abs-diff*.

Second, an insertion randomly selects a zero element in the tf-idf vector and inserts a new value for each year with probability $i$. The new value is drawn from the distribution of tf-idf values in the real network. An insertion is thus equivalent to adding a word into an article. We set the probability $i$ to approximate the change in words in an article over time. To calculate this probability, we first compute for each edge and its two nodes, (iv) the Manhattan distance between the two tf-idf vectors, which we call *man-dist* and is intuitively the number of different words used between the two articles. Interestingly, there is a strong and statistically significant correlation between (i) *year-diff* and (iv) *man-dist* (**Fig. S3F**). This correlation suggests that the longer the time between the discoveries of two neighboring nodes, the more different words are used in the two articles of each node. This relationship reflects the correlation between (i) *year-diff* and (ii) *sum-abs-diff* that was used to calculate probability of point mutation $p$ and points to the possibility of a slow-and-steady process underlying scientific research and discoveries.

Hence, we set the probability of insertion $i$ as half of the slope of the linear regression between *man-dist* and *year-diff*. We use half of this slope for $i$ because there is a third step for mutations that mirrors insertion: deletion. A deletion selects a randomly chosen non-zero element in the tf-idf vector and sets it to zero. Because a deletion of a tf-idf element is equivalent to removing one word in an article, just as insertion is equivalent to adding one word in the article, we set the probability of deletion $d$ to the probability of insertion $i$.

*Node creation.* In a real network, there is a distribution of cosine similarities between tf-idf vectors of neighboring nodes; the mean of that distribution is around 0.3 (**Fig. S4B**). In creating nodes from seeds in the simulation, we wished to match the distribution of cosine similarities in the real network. To do so, when we initialize a seed, we draw a value from a normal distribution with a mean and standard deviation of the cosine similarities of the real network. Once the cosine similarity between a seed and its parent becomes less than the drawn value, we add the seed as a node in the network.

*Node connection.* Once a node is added to a network, it must create connections to the rest of the network. To imitate the same process in a real Wikipedia article, we created a title of a new node, much like the title of a Wikipedia article. As the title, we selected ten words in each new node with the strongest tf-idf values excluding stop words. Then, if an existing node has a majority (i.e., six) of the words in the title, we create an edge from the new node to the existing node. The node will then be connected to the rest of the network, and in the next year of the simulation, the node will create a new *seed* on which to mutate.

Temporal modularity structure

To detect modules across time, we first built a multilayer network from our growing networks. In the multilayer network, each layer is a subset of a full network at a certain year. Each node in a layer (e.g., at time $t$) is connected to the equivalent node in the next layer (e.g., at time $t+1$) with weight 0.01. We empirically chose a low weight to detect changes in modularity when nodes are only being added and not removed. We then used the Python implementation of Ref. (*18*) in the software package *leidenalg* (version 0.8.1) to compute the modules to which each node belongs. We set the parameter *partition_type* to *ModularityVertexPartition* to partition based on modularity, *interslice_weight* to 0.01, and *n_iterations* to -1 to run iterations until there is no improvement in modularity. The structural stability signature (**Fig. 3C**) is robust to variations in the choice of the *interslice_weight* as long as the weight is small enough, i.e., on the order of $10^{-2}$ or less, to reveal changes in modularity structure (**Fig. S6**). At slight changes to the *interslice_weight*, only the magnitude of the signature changes without changes to the shape of the signature.

When we computed module membership, we observed that module memberships changed at different rates across time with almost no changes in module memberships in the second half of the time. We thus hypothesized that there may be "epochs" in the stability of module memberships, especially with a stabilization of module memberships later in history. To identify epochs in the number of changes in module membership, we first computed the number of changes in module membership across time as any time when the module membership is different than in the previous time point. Then, we summed the number of changes for each time point to obtain a single variable across time. Next, we used the R implementation of binary segmentation in the software package *changepoint* (*19*). We used the function *cpt.meanvar* to detect changes in both mean and variance of the signal, which is the number of changes in module membership over time. We set the parameter method to "BinSeg" for binary segmentation, Q to 3 for the maximum number of changepoints, and *test.stat* to "Poisson" for a Poisson distribution. We chose these settings because the PELT segmentation algorithm, which selects the optimal number of changepoints Q, selected a Q of 3 for 14 out of 28 networks, with a median and a mode of 3 (**Fig. S7**). Binary segmentation, on the other hand, allows the user to select a value for Q. So, we used binary segmentation with a Q of 3 for consistency across subjects. Additionally, we used a Poisson distribution because each change in module membership occurs independent of whether a node changes module membership in the previous time step. All other parameters were set to the default parameters. The algorithm then produces three indices, one for each changepoint, giving us four epochs (**Fig. 3B-C**).

Impulse response

To quantify the impact of an article-node on the subject network, we use the impulse response measure from linear systems theory (*28*). Here, the network is represented as an adjacency matrix $A$ such that an item $a_{ij}$ in the matrix, with row $i$ and column $j$, is the weight of the edge from the $j^{th}$ to the $i^{th}$ node. Then, the impulse response of node $i$ at time $m$ is given by the $i^{th}$ diagonal element of the controllability Gramian,

$$W_C = \sum_{m=0}^{K} A_{norm}^m BB^{\mathrm{T}} (A_{norm}^{\mathrm{T}})^m \,,$$

where $A_{norm}$ is normalized by dividing by one plus the dominant eigenvalue of $A$, and where $B$ is a vector of ones (*28*). Mathematically, this value quantifies the linearized response of the network to activity of a node $i$. The activity of the node and the subsequent network response can be intuited in the context of Wikipedia networks as a conceptual influence of node $i$ on the rest of the network.

In calculating impulse responses, we built a large network that consisted of all nodes in all subjects under study in order to quantify the influence of a node on all topics. In addition, we took the impulse response to a time horizon $m$ of 5 to capture up to five steps in the propagation of the impulse to the network. Because the dimensionality of the network is on the order of $10^5$, it is also computationally prohibitive to compute $A^m$ with larger values of $m$. Shorter time horizons also capture the relationship between impulse response and participation in the birth and death of cavities (**Fig. S8**).

Nobel prizes

We used Nobel prizes in Physics, Chemistry, and Physiology or Medicine as an external measure of influence. To identify which nodes in the concept networks received Nobel prizes, we parsed the Wikipedia articles "List of Nobel laureates in Physics", "List of Nobel laureates in Chemistry", and "List of Nobel laureates in Physiology or Medicine". In the section "Laureates" for each article, there is a table of Nobel laureates that includes a Rationale column, which describes the work of a laureate that motivated the Nobel prize with hyperlinks to articles that describe the laureate's discoveries. For example, for the scientist Maria Skłodowska-Curie, the Rationale column states, "for their joint researches on the radiation phenomena discovered by Professor Henri Becquerel" with a hyperlink to the article "Radiation". By obtaining all hyperlinks to articles in the Rationale columns, we identified nodes in the concept networks that were Nobel prize-winning nodes. All other nodes were identified as non-Nobel prize-winning.

**Supplementary Text**
Operationalizing philosophical theories of scientific progress

In the past century, philosophers of science have hypothesized about processes of growth in scientific knowledge. In this study, we have formulated the body of knowledge as a concept network (*29*) and have operationalized such hypothesized processes in terms of changes in the structure of the network during growth. Here, we briefly discuss our models of these hypotheses using a complex systems approach (*30*) and expand upon the insights revealed by our findings.

*Feyerabend*. While Feyerabend is chronologically the last of the philosophers whom we discuss, we tested his hypothesis first because his ideas about the growth of scientific knowledge are the most different from the rest. Feyerabend posits that there is no single set of scientific methodologies employed in practice by all scientists, in contrast to the Popperian and Khunian positions that Science does have a characteristic pattern of discovery (*4*). He was accepting of and even promoted a competition of theories and was careful to not demarcate science and its practice from other human endeavors, including storytelling and myth, to counter the effects of a history of power structures and the influence of Western philosophy upon, and in justification of, science. To model Feyerabend's theory, we use an edge-rewired network which produces a random network topology, similar to a Maslov-Sneppen random network (*27*). It is critical to note here that we are not modeling the scientific process that he describes as random rewiring of connections; instead, we are modeling the network structure resulting from science without a characteristic pattern of discovery as the culmination of random rewiring of connections. In comparisons between random, edge-rewired networks and their real counterparts, we observed that real networks often display clustering on various scales: at the scale of nodes with the clustering coefficient and at the mesoscale with modules and cores (*10–12*). Thus, a network formulation of knowledge suggests that the processes underlying network growth constrain the network topology.

Moreover, for Feyerabend, new significant discoveries in Science come as a result of reframing or seeing nearly all things completely anew, rather than filling-in otherwise well-recognized gaps in knowledge. We have shown that novel discoveries are the result of new cavities being formed or filled in the network and that these cavities are of higher-dimension, shorter period, and decreased frequency in real networks versus random or genetic ones. These results do not support the Feyerabend position of incommensurability and of novelty in knowledge being about a complete reconfiguration of the concept network.

*Lakatos*. Lakatos hypothesized that science progresses as a *research programme*, which has a common "hard core" of postulates with an auxiliary belt of hypotheses that builds upon the core (*3*, *31*, *32*). He was interested in building up a methodology of science (less so, an epistemology of science) whose focus is less on demarcation (what counts as Science or not) than on scientific practice. Within Lakatos' research programme, the "hard core" is a set of theories, practices, and commitments that most scientists would not want to give up in their research. Auxiliary hypotheses link the "hard core" to experiments and observations. Lakatos held that, in practice, science often chooses to modify auxiliary theories rather than give up on any of the "hard core" set of commitments, theories, and practices.

In this study, we operationalized Lakatos's hypothesis as growth within the core-periphery structure of a concept network (*12*). From the perspective of network topology, nodes in the "core" are densely connected to each other and form the topological center of the network; nodes in the "periphery" are loosely connected to the core. From the perspective of modifications to theories, it is more difficult to modify nodes in the core since they are densely connected to each other; moreover, it is easier to modify nodes in the periphery since they are only loosely connected to the network. The core-periphery structure hence reflects Lakatos's research programme with respect to both topology and scientific modification. Here, while we observed that concept networks do display a core-periphery structure, we also observed that concept networks grow "outward", with core nodes preceding neighboring peripheral nodes, only on average and that they often grow "inward", with core nodes preceded by neighboring peripheral nodes. This result supports an interesting aspect of Lakatos's theory of science: it is seen as a layered core with an outer layer that is often updated by new discoveries that are influenced, in turn, by discoveries that occur in the periphery.

*Kuhn*. Kuhn's ideas of scientific progress posit that there are two periods of science (*2*). One period is called *normal science* in which scientists "solve puzzles" within the current view of the body of knowledge, which he called a *paradigm*. The other period is a *paradigm shift* in which the current paradigm is overturned by another. In our study, we first operationalized "puzzle-solving" normal science as filling knowledge gaps, which we formulate as a cavity. Using persistent homology from the field of algebraic topology, we can identify cavities within a concept network and the times when they are created and destroyed throughout history (*16*). By comparing real networks to edge-rewired networks and our genetic model of knowledge growth, we observed that cavities in real networks are filled more quickly than in either the edge-rewired networks or genetic models. Moreover, real networks currently have fewer unfilled cavities and also have more cavities with higher dimensions. These results suggest that scientists create and fill knowledge gaps with scientific progress.

In addition to normal science operationalized as creating and filling cavities, we operationalized paradigms as the modularity structure of concept networks throughout history. Indeed, the view that scientists hold about a body of knowledge can depend on how a subject can be organized into its parts, or modules. For example, if a certain concept originally in the fringe of a large module engenders enough discoveries, the concept may start a new field of study and, at the same time, cause nodes to change their membership from the existing module to a new module. As another example, when an incommensurable paradigm shift occurs, one may expect that knowledge is reorganized into new, unrecognizable modules.

To detect changes in module membership across time, we formulate a growing concept network as a multilayer network with each layer containing the subnetwork that contains the node that were present at a particular year. When we detected module membership across time (*18*) and identified regimes of high or low change in module membership (*19*), we observed that concept networks went through different periods of varying durations and of varying intensity. In fact, we found a signature pattern of changing module membership across subjects: a short period of little to no change, then a long period of small but constant change, then a short burst of many changes, and finally, a long period of little to no change. Importantly, these changes in modularity structure do not result in completely new modules, as one may expect with the "incommensurability" of paradigm shifts but rather represent gradual changes to the modularity structure.

Interestingly, cavity-filling in real networks reveals the importance of curiosity as a catalyst for scientific progress. Kuhn recognized early on—see his early essay "The Essential Tension: Tradition and Innovation in Scientific Research" from 1959 (*33*)—that "normal science" would push scientists (and knowledge formation) into a constrained or conservative path—one that does not encourage true creativity and innovation. For Kuhn, the innovation occurred either when iconoclastic individuals were lucky enough to uncover something novel or via the accumulated failures of a scientific research programme that then made the search for more creative solutions necessary and better rewarded. While we provide no analysis on individuals, scientific commitments, or practices, our results on cavity-filling and its propensity for being rewarded in the scientific community suggest not two disparate periods of constrained normal science and innovative paradigm shifts but that scientists are continually and collectively driven, perhaps by curiosity, to uncover novel knowledge that "connects the dots" of existing knowledge.

Assumptions and limitations of network models of scientific knowledge

Our primary tool for testing philosophical theories of scientific progress is the formalization of a body of knowledge into a network of concepts and their inter-connections. This formalization allows the quantification of the structure of knowledge at different times in history and hence a data-intensive, statistical interpretation of philosophical theories (*34*). Here, we discuss the assumptions and limitations of network models of scientific knowledge.

*Models of minds and of reality.* In this study, we make the assumption that the growing Wikipedia networks model how *minds* have collectively built networks of knowledge over the course of history and codified that knowledge in hyperlinked wikis. After Pierre Duhem's criticisms of a simple Newtonian inductive method (i.e., the notion that science proceeds by generalizing from observations to theories) (*35*) and Karl Popper's "Critical Rationalism" and his method of "Falsification" to demarcate Science from other forms of knowledge (*36*), most Philosophers of Science in the second half of the 20th century turned to descriptions of how Science (sometimes science without a capital "s" in Feyerabend's case) works not only in theory but in practice. This turn sometimes included discovery and commitments to Realism—the notion that there is a Real world out there that Science uncovers—but not always (*37*). In various works, but in particular those of Kuhn and Lakatos, attention turned away from questions of Realism and more towards the *practice* of science and of scientific discovery. Hence, the growth of knowledge is cast in this light and in the terminology of theoretical propositions, theoretical commitments, "hard core" theses, "auxiliary" hypotheses and experimental observations to describe changes in the practice of scientific research over historical time. Thus, Reality sets the boundaries of what humans and their models can discover, but the two spaces are not coterminous. Correspondingly, one can make the further assumption that the mind, or collective minds as represented by a Wikipedia network, has been shaped by evolutionary forces to reflect the Real as closely as possible and lands upon one of several possible solutions in the space of the Real.

*Concepts versus commitments and practices.* A network of Wikipedia entries is different than a network of commitments, both theoretical and empirical, or of practices, which is often described by philosophers of science. Indeed, social and cultural values play an important role in the structuring of knowledge (*38*). Thus, we acknowledge that we formalize a body of knowledge only as its concepts and the interconnected relationships between the concepts without an account of theory, process, and observation. As such, the findings presented here would ideally be considered alongside qualitative studies of theory, process, and observation for a fuller picture of the structure of scientific knowledge. While Wikipedia networks themselves do not capture the discussions and theoretical commitments that are integral to scientific discovery, we add time as an additional dimension to our network analysis of concepts. Once we add time to networks, we can see changes in the structure of scientific knowledge over time, from which we may be able to quantitatively describe processes of scientific discovery.

*Concepts as nodes.* In a concept network, each node represents a concept and is named after the title of a Wikipedia article. While Wikipedia articles in the hard sciences are relatively accurate (*39*), we acknowledge some limitations of representing concepts as the titles of Wikipedia articles. First, without the proper context, the title of a Wikipedia article by itself can be ambiguous with respect to the concept to which it refers. Second, not every Wikipedia article is a scientific concept; a Wikipedia article can be about other topics, such as scientists or scientific books. Both of these limitations in the representation of concepts, however, are mitigated by our network formulation. By linking nodes according to their inter-dependence (see next paragraph), we provide a context for a concept-node with its neighboring concept-nodes. For example, the Wikipedia article on "On the Origin of Species" contains hyperlinks to "evolutionary biology", "biodiversity", "common descent", "tree of life", and "transmutation of species", which help disambiguate and further define concepts in networks. Moreover, corpus-based semantic analysis can provide description and even explanation of semantics based on context (*40*). Lastly, while concepts themselves may change over the course of history, we use Wikipedia articles archived at a single time in their history, which runs this risk of obscuring how concepts slowly evolve to form new concepts (**Fig. S3B**). As the work of conceptual history (*Begriffsgeschichte*) suggests, historical semantics—understanding where terms have come from and how their meanings have changed over time—is not only important for historical reasons, it also conditions contemporary thought and practice (*41, 42*). Future study would ideally situate the concept-nodes examined here in cultural-linguistic context over time.

*Concept relationships as edges.* We formulate the relationship between concepts as the hyperlink between two Wikipedia articles. To illustrate the intuition for this formulation, suppose that there are two Wikipedia articles *A* and *B*. Article *A* hyperlinks to article *B* when article *A* uses in its description the word or concept that is described in article *B*. Thus, article *B* influences, or even in some case is necessary for, the description of article *A*, and we add an edge from article *B* to article *A*. Hence, a network of hyperlinked articles represents a network of *influence* or *dependence* between concepts. We acknowledge, however, that hyperlinks involve some aspect of chance, i.e., the willingness of an editor to write a linking content page and to hyperlink it. To ensure that only that most relevant hyperlinks are captured in the network models, we only use hyperlinks in the "lead" introductory section of each article.

*Network structure versus process.* The network structure is different from the process used to obtain that structure. In this study, we aim to infer the process that produced the concept network that we see today in the structure of Wikipedia. In fact, one may expect multiple processes to result in the same network structure. We therefore make the assumption that a resultant network structure sets the boundaries to the types of processes that have built the network.

Statistical tests for core-periphery lead-lag

As a supplement to our analysis in **Fig. 1C**, we performed two-sided one-sample t-tests for for the lead-lag values for all core-periphery edges in each network with a null hypothesis that the mean is 0, where the lead-lag value is the year of a core node minus the year of each of its neighboring peripheral nodes. **Table S3** shows the t-statistic and p-values for t-tests for all subject-networks. A negative t-statistic indicates that core nodes are, on average, born before each of their neighboring peripheral node. Not all t-statistics are negative, and for subjects with negative t-statistics, the maximum p-value is 0.21. Thus, we cannot statistically conclude that in all subjects, core nodes on average precede their neighboring peripheral nodes.

Robustness of core-periphery lead-lag

To test for robustness of our core-periphery lead-lag analysis, we measured the core-periphery lead-lag (i) within modules, (ii) for all subjects, and (iii) across epochs. First, we tested whether knowledge indeed grows outward but within modules, in which case, the core-periphery lead-lag for an entire network would not reveal that core nodes lead peripheral nodes. By maximizing modularity (*43*) and core-periphery (*44*), we can segregate nodes by modules and core-periphery. So, we identified modules in each subject-network using the Clauset-Newman-Moore greedy modularity maximization (*11*). Within each module of each subject-network, we identified the core and periphery nodes using the Borgatti-Everett core-periphery detection algorithm (*12*). Then, we computed the core-periphery lead-lag as the year of core nodes minus the year of neighboring peripheral nodes for each core-periphery edge. In accordance with our results in **Fig. 1C**, we found that core nodes do not necessarily precede peripheral nodes within modules (**Fig. S1A**).

Second, we computed the core-periphery lead-lag for all core-periphery edges in all networks to test whether core nodes precede peripheral nodes on average across all subjects (**Fig. S1B**). We performed a two-sided t-test for the lead-lag values for all core-periphery edges in all networks with a null hypothesis that the mean is 0. We obtained a t-statistic of -58.9 ($p \ll 0.001$). Thus, on average for nodes of all subjects, core nodes precede the periphery but clearly not always. As we noted in the main text, this holistic statistical test supports a nuanced variation to Lakatos's original hypothesis: that while the body of scientific knowledge does not have a "hard" core from which knowledge grows strictly outward, knowledge tends on average to grow outward from a "soft", malleable core.

Third, we tested core-periphery lead-lag in a time-dependent way to ensure robustness of our results to differences in core-periphery structure at earlier time points in a network's growth. Thus, for each subject-network, we created subnetworks from the nodes that were present at ten time points in its history. The ten time points, or epochs, were equally spaced such that there is an equal number of unique years of nodes in between each time point. As illustrated in three example subjects (**Fig. S1C**), the t-statistic becomes more negative across epochs, suggesting that core nodes do not always precede peripheral nodes even with core-periphery structures of a network at earlier time points in its growth. Taken together, these supplementary analyses demonstrate the robustness of our core-periphery growth analysis within modules, for all subjects, and across epochs.

Robustness of results to slight changes in years of nodes

Our analyses of cavity-filling and the paradigm shift signature are sensitive to the discovery-years of the nodes. Therefore, to ensure the robustness of our results to slight changes in the estimated year of nodes, which could be due to any systematic or random errors, we performed our analyses for the paradigm shift signature and persistent homology again with networks that have years "jittered" by plus or minus one year. Assessing robustness in this way is important especially given that not all Wikipedia articles have a history section or a year of discovery (**Fig. S9A**). To jitter the years of real networks, we added a -1, 0, or 1, drawn uniformly with each number with a probability of $1/3$, to the year of each node in all networks. We observed that the shape of the paradigm signature is robust to jittering in both the magnitude and duration (**Fig. S9B**).

In addition, we observed that our cavity-filling results are robust to jittering. The duration of cavities is slightly lower in jittered networks than in real networks ($KS = 0.04, p = 2.1 \times 10^{-4}$; **Fig. S9C**). We note that this difference exists for short lifetimes on the order of 10, which is in contrast to **Fig. 2D** in which the differences are present for longer lifetimes on the order of 100 and 1,000. For the cavities that are currently present and for the dimensions of cavities, the cumulative frequencies are the same in the jittered networks as in the real networks (**Fig. S9D-E**). These supplementary analyses demonstrate the robustness of our results on cavity-filling and the paradigm shift signature to slight jittering of the years of nodes.

Diversity statement

Recent work in several fields of science has identified a bias in citation practices such that papers from women and other minority scholars are under-cited relative to the number of such papers in the field (*45–48*). Here, we sought to proactively consider choosing references that reflect the diversity of the field in thought, form of contribution, gender, race, ethnicity, and other factors. First, we obtained the predicted gender of the first and last author of each reference by using databases that store the probability of a first name being carried by a woman (5, 6). By this measure (and excluding self-citations to the first and last authors of our current paper), our references contain 25.53% woman(first)/woman(last), 9.09% man/woman, 10.64% woman/man, and 54.74% man/man. This method is limited in that a) names, pronouns, and social media profiles used to construct the databases may not, in every case, be indicative of gender identity and b) it cannot account for intersex, non-binary, or transgender people. Second, we obtained predicted racial/ethnic category of the first and last author of each reference by databases that store the probability of a first and last name being carried by an author of color (*49, 50*). By this measure (and excluding self-citations), our references contain 5.06% author of color (first)/author of color(last), 15.64% white author/author of color, 16.35% author of color/white author, and 62.95% white author/white author. This method is limited in that a) names, Census entries, and Wikipedia profiles used to make the predictions may not be indicative of racial/ethnic identity, and b) it cannot account for Indigenous and mixed-race authors, or those who may face differential biases due to the ambiguous racialization or ethnicization of their names. We look forward to future work that could help us to better understand how to support equitable practices in science.

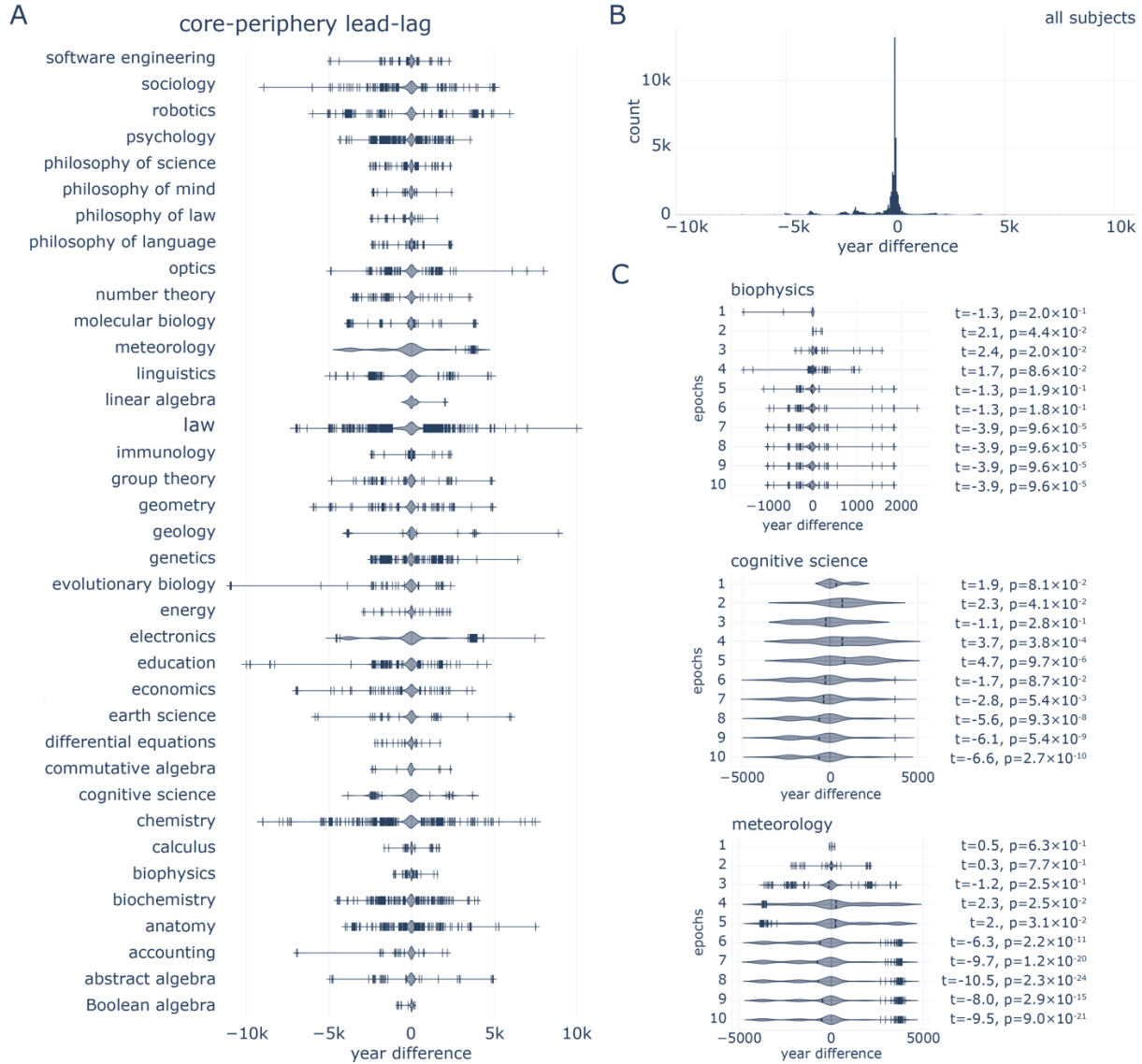

**Fig. S1. Core-periphery lead-lag relationship within modules, for all subjects, or across epochs.** (A) Core nodes do not necessarily precede their neighboring peripheral nodes within modules. Violin plots show distributions of the year of the core nodes minus the year of its neighboring peripheral nodes, for each core-periphery edge. Negative values indicate that core nodes are discovered before their neighboring peripheral nodes, and *vice versa*. Vertical lines indicate outliers, which are either less than $Q_1 - (1.5 \times IQR)$ or greater than $Q_3 + (1.5 \times IQR)$ where $Q_n$ is the $n^{th}$ quartile and $IQR = (Q_3 - Q_1)$. (B) Distribution of core-periphery lead-lag for all core-periphery edges in all subjects. While core nodes do not always precede their neighboring peripheral nodes, core nodes do so on average ($t = -58.9, p = 0$; null hypothesis that sample mean is 0). (C) Core-periphery lead-lag plots computed for subnetworks at ten epochs in the history of each network for three example networks. Cores do not always precede their neighboring peripheral nodes even when core-periphery is calculated at previous time points in the history of a network.

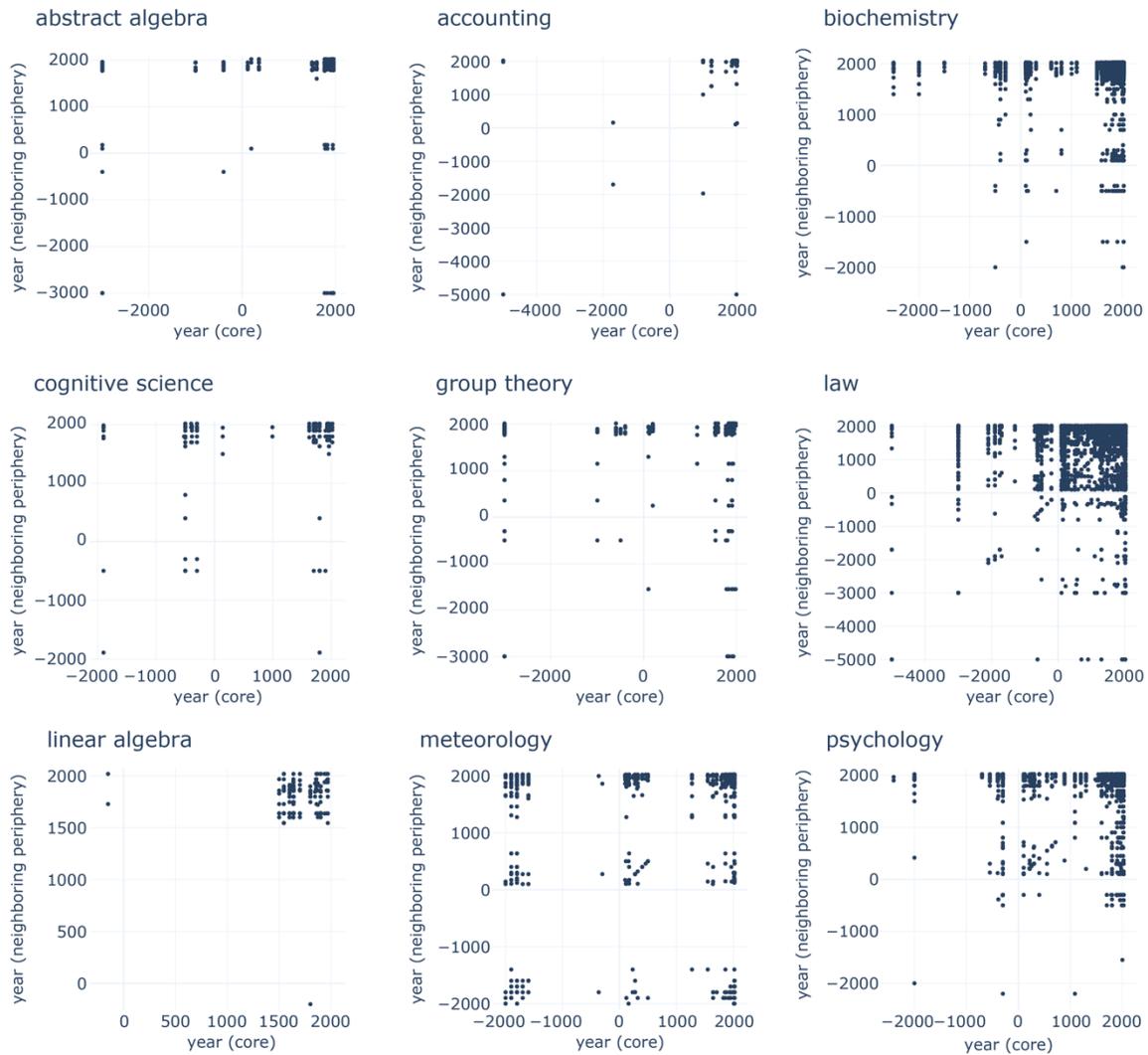

**Fig. S2. Some core nodes are born early while other core nodes are born after their neighboring peripheral nodes.** While most core nodes both precede and follow their neighboring peripheral nodes, some core nodes are consistently born before neighboring peripheral nodes. The plots show the years for nodes in the core (x-axis) against the years for neighboring nodes in the periphery (y-axis) for nine example subjects. For each plot, a point on the top-left side indicates that a node in the core was born before a neighboring periphery, whereas a point on the bottom-right side indicates that a node in the core was born after a neighboring periphery. The peripheral nodes that are neighboring most core nodes are born both before and after the births of their neighboring core nodes. Some core nodes (points on the very top left) are born before most, if not all, of their neighboring peripheries.

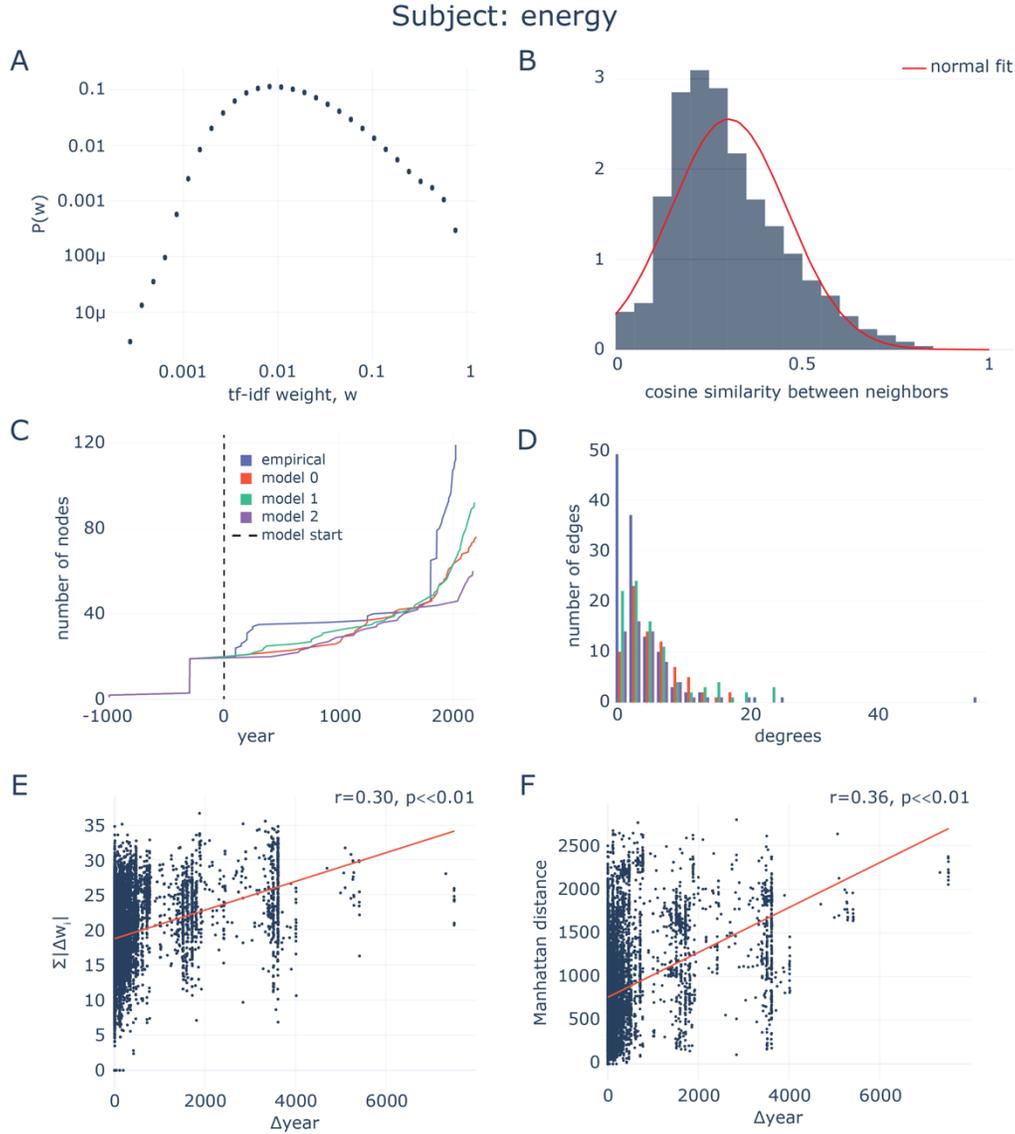

**Fig. S3. Statistics of tf-idf vectors in an example real network.** For the example subject "Anatomy", the statistics of the network used to inform simulations of knowledge discovery. (A) Distribution of tf-idf weights on a log-log plot. (B) Distribution of cosine similarity of tf-idf vectors between neighbors. (C) In three simulations, the model follows the growth of the real network and shows an exponential increase in the number of nodes. (D) The degree distributions in models are similar to that of the real network. (E) The sum of the absolute difference in tf-idf values between neighbors plotted against the year difference between neighbors. The correlation between the quantities suggests that more distant knowledge takes longer to discover. The vertical striations are due to temporal clusters of discovery; for example, in panel C, we can see that there are bouts of discoveries around the 3rd century BC and the 2nd and 17th centuries AD. Red line indicates line of best fit. (F) The Manhattan distance between tf-idf vectors between neighbors plotted against the year difference between neighbors.

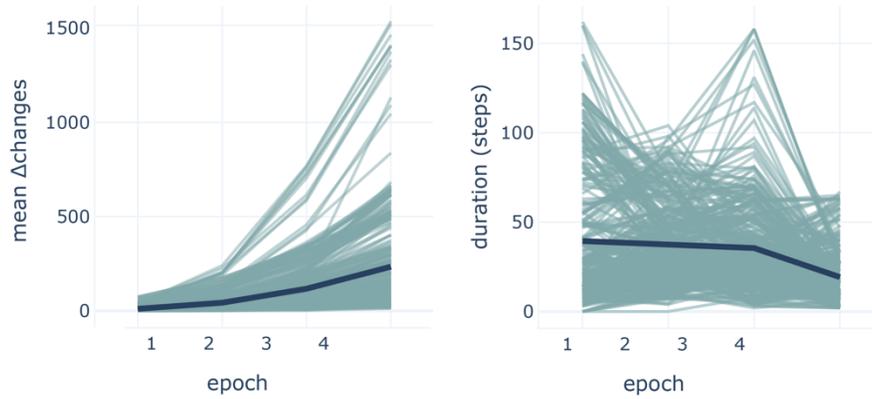

**Fig. S4. Paradigm shift signature breaks in edge-rewired networks.** The mean number of changes within an epoch (left panel) and the duration of each epoch in time steps (right panel) reveals a different signature (dark green) in paradigm shifts across subjects (teal) in edge-rewired networks than that observed in the true data (see for comparison **Fig. 3C**).

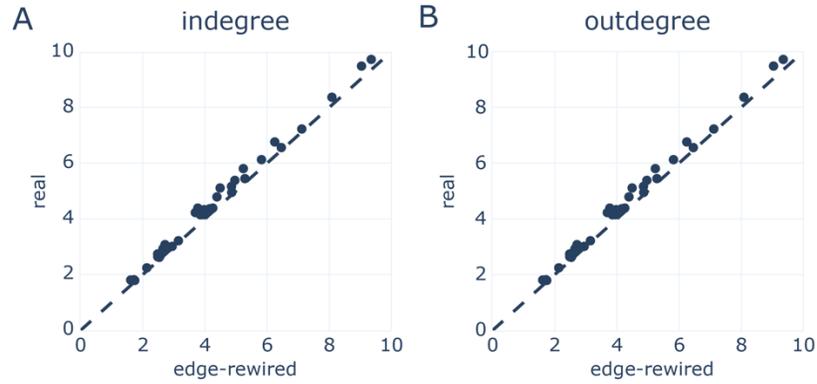

**Fig. S5. Degree distributions of edge-rewired null networks.** (A-B) The average indegree (panel A) and average outdegree (panel B) of real concept networks compared to their edge-rewired nulls. Note that the average indegree and outdegree are identical because for any network, the total indegree is equivalent to the total outdegrees.

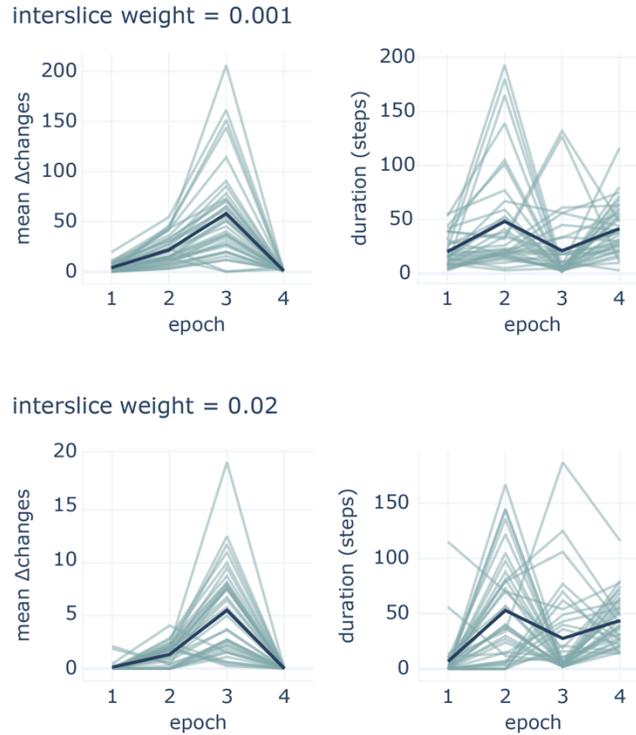

**Fig. S6. The structural stability signature observed in Fig. 3 is robust to changes in the interslice weight.** The mean number of changes within an epoch (left panels) and the duration of each epoch (right panels) reveals a signature in paradigm shifts (dark green) averaged across subjects (teal) for interslice weights 0.001 (top panels) and 0.02 (bottom panels). For different interslice weights (top, bottom, and in **Fig. 3C**), the magnitude of the signature changes without changes to the shape of the signature.

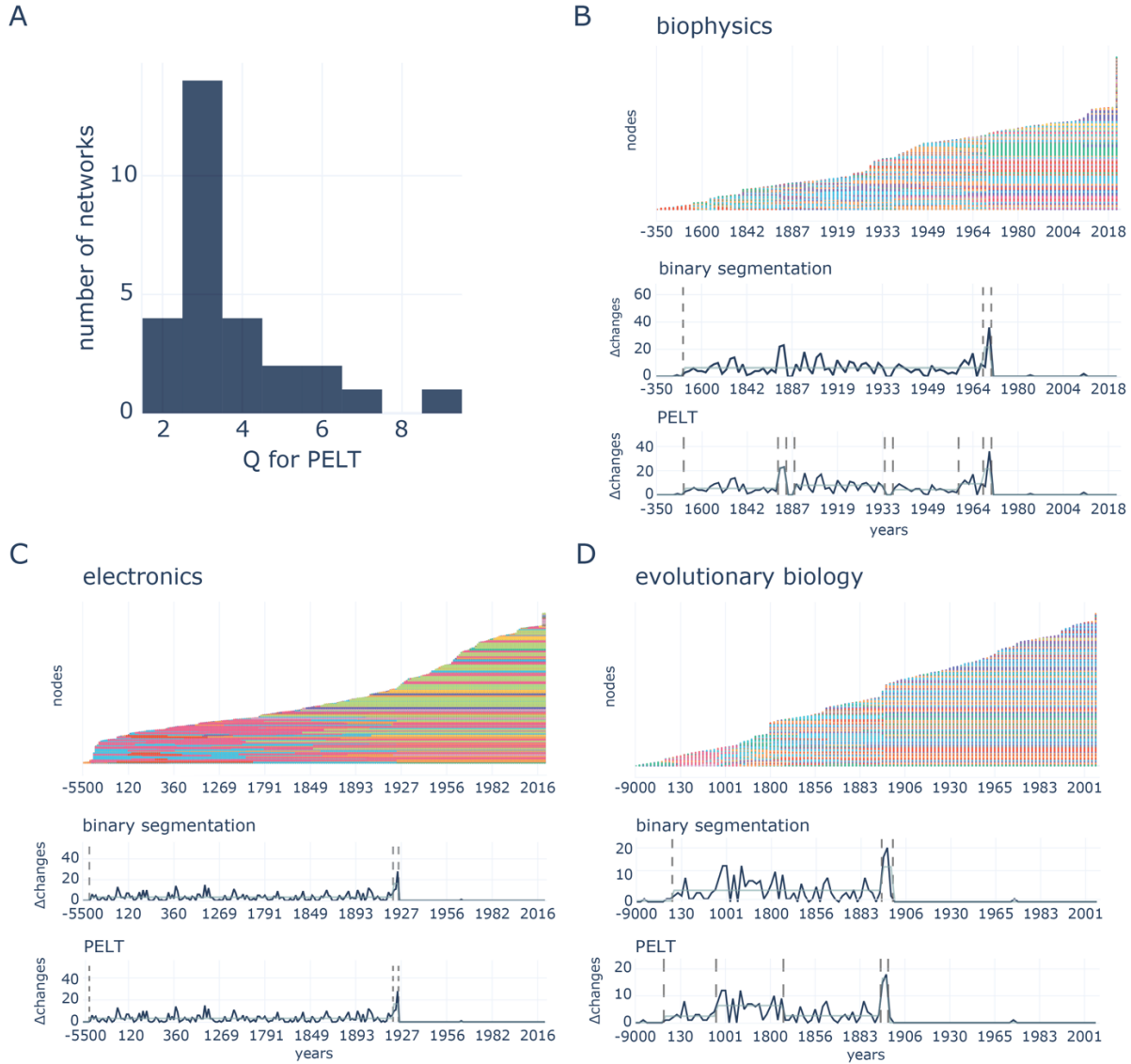

**Fig. S7. Comparison of changepoint detection using PELT versus binary segmentation.** (A) Distribution of optimal number of changepoints Q discovered by the PELT algorithm for concept networks. The median and mode is 3. (B-D) Comparison of changepoint detection using PELT versus binary segmentation with a Q of 3 for the networks *biophysics* (panel B), *electronics* (panel C), and *evolutionary biology* (panel D). For each of the panels B, C, and D, the top plot shows the module membership across time points, and the bottom two plots show the number of changes in module membership with the grey dashed lines indicating changepoints detected by either binary segmentation or PELT and the light green line indicating the average number of changes between each pair of changepoints. Most changepoints detected by binary segmentation with Q of 3 matches the changepoints detected by PELT with an optimal Q, as in the *electronics* network in panel C.

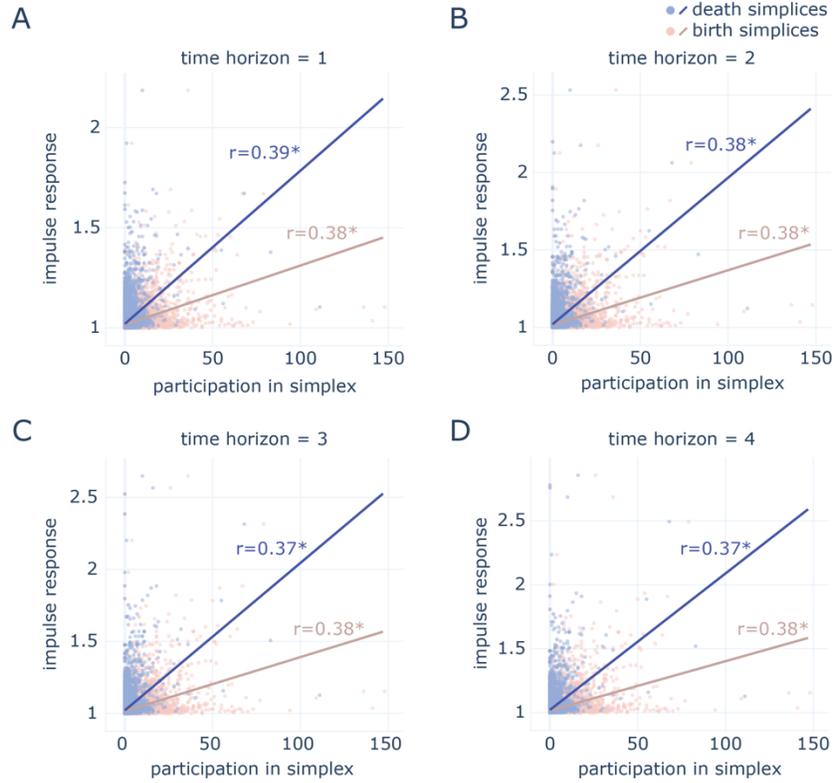

**Fig. S8. The shorter time horizons capture the relationship between impulse response and participation in the birth and death of cavities.** (A-D) The correlation between impulse response of nodes and the frequency of participation of nodes in the birth and death of cavities is maintained across shorter time horizons (all $p \ll 0.001$). A small decay in correlation, from around 0.38 at a time horizon of 1 to 0.36 at a time horizon of 5, demonstrates a robustness in the relationship between the cavity participation and the impulse response of nodes.

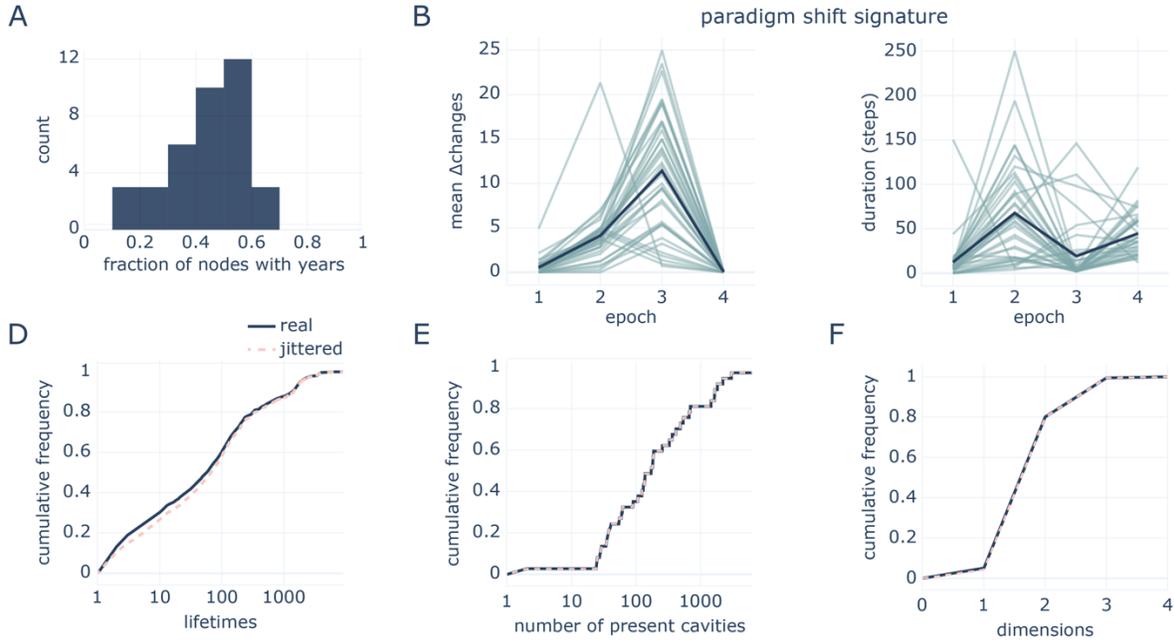

**Fig. S9. Paradigm shift signature and cavity-filling statistics are robust to slight changes in the year of nodes.** (A) Distribution of the fraction of nodes whose articles have a year of discovery across subjects. (B) Signature in paradigm shifts (dark green) averaged across subjects (teal) in the mean number of changes within an epoch (left panel) and the duration of each epoch (right panel) for jittered networks. (C) Duration of knowledge gaps is slightly lower in jittered networks than in real networks ($KS = 0.04$, $p = 2.1 \times 10^{-4}$). We note that this difference exists for short lifetimes on the order of 10, which is in contrast to **Fig. 2D** in which the differences are present for longer lifetimes on the order of 100 and 1,000. (D) The frequency of knowledge gaps that are currently present (i.e., that have yet to die) is the same in jittered networks as in real networks. (E) The frequency of cavity dimensions is the same in jittered networks as in real networks.

**Table S1. Network measures.**

| Measure | Code | Equation | Reference |
|---|---|---|---|
| Clustering | `networkx.clustering()` | $$c_n = \frac{2T(n)}{k_n(k_n - 1)}$$ where $T(n)$ is the number of triangles through node $n$ and deg($n$) is the degree of node $n$ | (10) |
| Modularity | `networkx.algorithms.community.modularity_max.greedy_modularity_communities()` | $$Q = \frac{1}{2m}\sum_{ij}\left[A_{ij} - \frac{k_i k_j}{2m}\right]\delta(g_i, g_j)$$ where $m$ is the number of edges in a network, $A$ is the adjacency matrix of the network, $i$ and $j$ are nodes, $g_i$ is the index of the community (or module) to which $i$ belongs, and $\delta(g_i, g_j)$ is 1 if $g_i$ and $g_j$ are equal and 0 otherwise. | (11) |
| Coreness | `bct.core_periphery_dir()` | $$\rho = \sum_{ij} a_{ij}\delta_{ij}$$ where $a_{ij}$ is the connection from node $i$ to node $j$ and $\delta_{ij} = 1$ if $c_i$ = CORE or $c_j$ = CORE and 0 otherwise. | (12) |
| Temporal modularity | `leidenalg.find_partition_temporal()` | $$Q = \frac{1}{2\mu}\sum_{ijsr}\left[\left(A_{ijs} - \gamma_s \frac{k_{is}k_{js}}{2m_s}\right)\delta_{sr} + \delta_{ij}C_{jsr}\right]\delta(g_{is}, g_{jr})$$ where for nodes $i$ and $j$ in slices $s$ and $r$, $2\mu = \sum_{jr}\kappa_{jr}$, $\kappa_{jr} = k_{js}c_{jr}$, $k_{js} = \sum_{is}A_{ijs}$, $c_{js} = \sum_r C_{jrs}$, $A_{ijs}$ is the weight of the edge from $i$ to $j$ in $s$, $\gamma_s$ is the resolution of slice $s$, $m_s = \sum_j k_{js}$, $\delta(g_i, g_j)$ is 1 if $g_i$ and $g_j$ are equal and 0 otherwise. | (18) |
| Changepoint detection | `cpt.meanvar()` | $$ML(\tau_1) = \log p(y_{1:\tau_1}|\hat{\theta}_1) + \log p(y_{(\tau_1+1):n}|\hat{\theta}_2)$$ where ML is the maximum likelihood for a given change point at $\tau_1$, $y_{1:\tau_1}$ is the signal from time 1 to $\tau_1$ $\hat{\theta}_1$ is the maximum likelihood estimate of parameters, in our case the mean and variance. | (19) |

**Table S2. Network metrics for subjects.** *N* is the number of nodes. Errors are standard deviations.

| Subject | N | Clustering | Modularity | Coreness |
|---|---|---|---|---|
| Boolean algebra | 77 | 0.15±0.145 | 0.34 | 0.69 |
| abstract algebra | 356 | 0.18±0.116 | 0.36 | 0.73 |
| accounting | 115 | 0.14±0.158 | 0.44 | 0.68 |
| anatomy | 2043 | 0.12±0.104 | 0.56 | 0.72 |
| biochemistry | 1061 | 0.11±0.097 | 0.35 | 0.79 |
| biophysics | 463 | 0.10±0.161 | 0.59 | 0.94 |
| calculus | 103 | 0.19±0.153 | 0.38 | 0.67 |
| chemistry | 1032 | 0.13±0.100 | 0.26 | 0.82 |
| cognitive science | 115 | 0.22±0.148 | 0.25 | 0.72 |
| commutative algebra | 88 | 0.17±0.125 | 0.22 | 0.70 |
| dynamical systems and differential equations | 144 | 0.14±0.171 | 0.58 | 0.71 |
| earth science | 103 | 0.25±0.144 | 0.33 | 0.69 |
| economics | 511 | 0.11±0.118 | 0.42 | 0.72 |
| education | 668 | 0.10±0.148 | 0.53 | 0.91 |
| electronics | 1145 | 0.10±0.114 | 0.44 | 0.85 |
| energy | 119 | 0.11±0.156 | 0.44 | 0.83 |
| evolutionary biology | 265 | 0.16±0.118 | 0.26 | 0.78 |
| genetics | 1111 | 0.12±0.104 | 0.32 | 0.87 |
| geology | 92 | 0.14±0.144 | 0.31 | 0.75 |
| geometry | 294 | 0.17±0.129 | 0.38 | 0.75 |
| group theory | 295 | 0.18±0.129 | 0.29 | 0.76 |
| immunology | 410 | 0.14±0.164 | 0.45 | 0.89 |
| law | 3174 | 0.08±0.106 | 0.43 | 0.86 |
| linear algebra | 138 | 0.22±0.149 | 0.32 | 0.74 |
| linguistics | 395 | 0.16±0.119 | 0.35 | 0.76 |
| meteorology | 626 | 0.13±0.138 | 0.44 | 0.81 |
| molecular biology | 395 | 0.15±0.119 | 0.27 | 0.77 |
| number theory | 276 | 0.17±0.159 | 0.48 | 0.75 |
| optics | 352 | 0.16±0.135 | 0.31 | 0.79 |
| philosophy of language | 235 | 0.15±0.169 | 0.44 | 0.85 |
| philosophy of law | 146 | 0.09±0.149 | 0.47 | 0.81 |
| philosophy of mind | 106 | 0.19±0.161 | 0.28 | 0.75 |
| philosophy of science | 357 | 0.14±0.161 | 0.45 | 0.84 |
| psychology | 1568 | 0.09±0.123 | 0.45 | 0.86 |
| robotics | 1099 | 0.11±0.157 | 0.56 | 0.86 |
| sociology | 630 | 0.07±0.112 | 0.46 | 0.77 |
| software engineering | 226 | 0.14±0.132 | 0.36 | 0.71 |

**Table S3. Core-periphery lead-lag t-tests for all subjects.**

| Subject | t-statistic | p-value |
|---|---|---|
| Boolean algebra | -1.73 | $4.4 \times 10^{-2}$ |
| abstract algebra | -11.70 | $4.0 \times 10^{-30}$ |
| accounting | -0.91 | $1.8 \times 10^{-1}$ |
| anatomy | -15.34 | $1.7 \times 10^{-52}$ |
| biochemistry | -15.30 | $1.1 \times 10^{-51}$ |
| biophysics | -3.94 | $4.8 \times 10^{-5}$ |
| calculus | -2.46 | $7.4 \times 10^{-3}$ |
| chemistry | -22.60 | $7.3 \times 10^{-107}$ |
| cognitive science | -6.62 | $1.4 \times 10^{-10}$ |
| commutative algebra | -3.03 | $1.4 \times 10^{-3}$ |
| dynamical systems and differential equations | -2.70 | $3.9 \times 10^{-3}$ |
| earth science | -0.82 | $2.1 \times 10^{-1}$ |
| economics | -4.94 | $4.6 \times 10^{-7}$ |
| education | -7.64 | $2.8 \times 10^{-14}$ |
| electronics | -19.83 | $5.6 \times 10^{-82}$ |
| energy | -2.40 | $8.9 \times 10^{-3}$ |
| evolutionary biology | 2.64 | $4.3 \times 10^{-3}$ |
| genetics | -10.58 | $3.9 \times 10^{-26}$ |
| geology | -1.14 | $1.3 \times 10^{-1}$ |
| geometry | -10.81 | $1.8 \times 10^{-25}$ |
| group theory | -11.18 | $3.8 \times 10^{-27}$ |
| immunology | -6.16 | $7.0 \times 10^{-10}$ |
| law | -27.16 | $2.9 \times 10^{-156}$ |
| linear algebra | -1.87 | $3.2 \times 10^{-2}$ |
| linguistics | -13.01 | $2.2 \times 10^{-36}$ |
| meteorology | -9.44 | $9.4 \times 10^{-21}$ |
| molecular biology | -1.67 | $4.7 \times 10^{-2}$ |
| number theory | -8.17 | $2.3 \times 10^{-15}$ |
| optics | -8.26 | $2.8 \times 10^{-16}$ |
| philosophy of language | -2.94 | $1.8 \times 10^{-3}$ |
| philosophy of law | -3.75 | $1.5 \times 10^{-4}$ |
| philosophy of mind | -7.20 | $1.1 \times 10^{-11}$ |
| philosophy of science | -4.43 | $5.7 \times 10^{-6}$ |
| psychology | -16.58 | $1.1 \times 10^{-59}$ |
| robotics | -18.37 | $6.8 \times 10^{-68}$ |
| sociology | -5.93 | $2.1 \times 10^{-9}$ |
| software engineering | -3.53 | $2.3 \times 10^{-4}$ |